\documentclass[aps,prl,letterpaper,superscriptaddress,longbibliography,twocolumn,asmath,amssymb]{revtex4-2}

\usepackage{graphicx, amsmath, subfigure}
\usepackage{amsbsy}
\usepackage{amsfonts}
\usepackage{hyperref}
\usepackage{braket}
\usepackage{color}
\usepackage{float}
\usepackage{chngcntr}
\usepackage{mathrsfs}
\usepackage{textcomp}
\usepackage{booktabs}      
\usepackage{dcolumn}       
\usepackage{siunitx}      
\usepackage{float}         

\begin{document}
	\preprint{}
	\title{Fermion-mediated Casimir effect on mesoscopic rings implementing non-Clifford $\mathbf{SWAP}^\alpha$ gates}
	
	\author{Liang Du}
	\affiliation{Interdisciplinary Center for Theoretical Physics and Information Sciences, Fudan University, Shanghai 200433, China}
	
	\author{Qing-Dong Jiang}
	\thanks{Corresponding author: qingdong.jiang@sjtu.edu.cn}
	\affiliation{Tsung-Dao Lee Institute \& School of Physics and Astronomy, Shanghai Jiao Tong University, Shanghai 200240, China}
	\affiliation{Hefei National Laboratory, Hefei 230088, China}
	
	\author{Yijia Wu}
	\thanks{Corresponding author: yijiawu@fudan.edu.cn}
	\affiliation{Interdisciplinary Center for Theoretical Physics and Information Sciences, Fudan University, Shanghai 200433, China}
	\affiliation{State Key Laboratory of Surface Physics, Fudan University, Shanghai 200433, China}
	\affiliation{Hefei National Laboratory, Hefei 230088, China}
	\date{\today}

\begin{abstract}
The Casimir effect is typically governed by intrinsic material properties and lacks in situ tunability. We show that, in mesoscopic rings, both the magnitude and sign of the fermion-mediated Casimir interaction can be controlled via the Aharonov–Bohm effect.
The resulting interplay between the Aharonov–Bohm phase and the Casimir interaction provides a route to engineer long-range interactions. In particular, this mechanism enables the implementation of non-Clifford $\mathrm{SWAP}^\alpha$ gates between spatially separated spin qubits, thereby reducing the overhead for universal quantum computation and quantum error correction in spin-qubit architectures.
\end{abstract}

\maketitle

\textit{Introduction.}---The Casimir effect was first introduced as a force between conducting plates due to electromagnetic vacuum fluctuations \cite{casimir1948attraction}. Its basic idea extends more broadly: quantum fluctuations of fermionic \cite{Fermion_Casimir1,bulgac2001casimir,sundberg2004casimir,beenakker2024topologically,flachi2017sign}, phononic \cite{schecter2014phonon,pavlov2018phonon,rodriguez2026phononic}, and other fields can also mediate effective forces between objects. Fermionic mediators are especially interesting because, unlike photons or neutral phonons, they carry charge and can therefore be controlled by electromagnetic fields. This extra handle, however, comes with a limitation. In most condensed-matter systems, electronic phase coherence extends only over relatively short distances, limiting the range of the resulting fermionic Casimir interaction. Mesoscopic rings notably provide an important exception \cite{buttiker1983josephson, PhysRevLett.62.587}. When the coherence length exceeds the ring circumference, electrons remain phase coherent around the entire ring and support persistent currents \cite{PhysRevLett.64.2074, PhysRevLett.67.3578, persistent_current}. They thus offer a natural setting for tunable fermionic Casimir physics.

In this Letter, we investigate fermion-mediated Casimir interactions in a mesoscopic ring geometry. Because electrons carry charge, their quantum spectrum can be tuned directly by electromagnetic fields. We show that both the sign and the magnitude of the Casimir interaction are controlled by the Aharonov--Bohm (AB) phase \cite{AB_effect} generated by a magnetic flux threading the ring. The sign controlling reflects the physics that AB flux violate the chiral symmetry and therefore the fermionic Casimir sign theorem {must be modified accordingly} \cite{Fermion_Casimir1}.
Furthermore, in contrast to $d$-dimensional bulk systems, where the Casimir interaction typically decays with distance $l$ as a power law $\sim l^{-d}$, we demonstrate that the Casimir effect in one-dimensional (1D) mesoscopic rings gives rise to a long-range interaction scaling as $\sim l^{-1}$. This tunable long-range interaction provides a powerful mechanism for engineering correlations between spatially separated quantum states.

An ideal platform to exploit such a tunable long-range Casimir effect is provided by spin qubits in semiconductor quantum dots (QDs). Spin-based quantum computation employs electron spins confined in QDs as qubits \cite{DanielLoss_spin_qubit, spin_qubit_PhysicsToday}.
%The concept of spin-based quantum computation  was introduced by Loss and DiVincenzo \cite{DanielLoss_spin_qubit, spin_qubit_PhysicsToday}, where electron spins confined in QDs encode qubits. 
The Heisenberg exchange interaction $-J \mathbf{S}_A \cdot \mathbf{S}_B$ between two spins $\mathbf{S}_A$ and $\mathbf{S}_B$ naturally implements a non-Clifford $\mathrm{SWAP}^\alpha$ gate (most notably the $\sqrt{\mathrm{SWAP}}$ gate) \cite{sqrtSWAP_universal_quantum_gate_1, sqrtSWAP_universal_quantum_gate_2} generating entanglement between two spin qubits, which together with single-qubit operations forms a universal gate set for quantum computation \cite{PhysRevLett.89.147902, sqrtSWAP_universal_quantum_gate_3}. Experimentally, recent advances have brought spin qubits to industry-compatible platforms while maintaining high fidelities \cite{industry_spin_qubit_2025, bartee2025spin}. Nonetheless, to date, two-qubit gates are typically realized via pulsed exchange interactions \cite{two_qubit_logic_gate_2015}, which decay exponentially with distance as $J \sim e^{-l/l_0}$, thereby restricting operations to nearest-neighbor qubits.

Implementing two-qubit gates between distant qubits (i.e., beyond nearest-neighbor couplings) can substantially reduce the overhead required for quantum error correction \cite{Surface_code_threshold, cohen2022low, PRXQuantum.6.020331}. In particular, long-range connectivity enables quantum low-density parity-check codes with a constant encoding rate (the ratio of logical to physical qubits) \cite{Gottesman_2014_constant_overhead, PRXQuantum.2.040101}, as demonstrated in platforms such as Rydberg-atom qubits \cite{RevModPhys.82.2313, PhysRevA.92.030303, Lukin, xu2024constant}. By contrast, the  surface code \cite{Kitaev_surface_code, surface_code_vanishing_encoding_Rate} with nearest-neighbor connectivity exhibits a vanishing encoding rate.
In the spin-qubit architectures, several approaches, including superexchange interactions between remote QDs, have been explored to realize long-range gates \cite{HQXu_InAs, HQXu_superexchange, mills2019shuttling, borjans2020resonant}. In this Letter, we propose an experimentally feasible scheme that exploits the tunable long-range fermion-mediated Casimir effect to implement non-Clifford $\mathrm{SWAP}^\alpha$ gates between spatially separated QDs in a controllable manner. Our proposal provides a new route toward overcoming the nearest-neighbor limitation of conventional exchange-based spin-qubit architectures.

% The theorem on quantum computing shows that a universal set of quantum gates on qubit requires the Clifford gate and at least one non-Clifford gates. For instance, if the non-Clifford gate is a SWAP gate that exchange the qubit status as $\ket{\beta_A \alpha_B} = \mathrm{SWAP} \ket{\alpha_A \beta_B}$. Such a SWAP gate, combined with the set of Clifford gate, as well as the Bell state, can be used to adopt a universal set for quantum computing. 
% This non-Clifford $\mathrm{SWAP}^\alpha$ gate can be implemented by two qubits on a mesoscopic rings. The  $\mathrm{SWAP}^\alpha$ gate operation is accomplished via the ABC effect between these two qubits via the fermion-mediated Casimir effect on this ring.

%%%%%%%%%% THIS IS A CUTTING LINE %%%%%%%%%%

\textit{Fermion-mediated Casimir effect on mesoscopic rings.}---We begin by revisiting the general formalism of the fermion-mediated Casimir effect \cite{Fermion_Casimir1}. We consider a fermionic field $(\Psi^\dagger, \Psi)$ coupled to two objects $A$ and $B$, described by local potentials $\hat{\mathcal{V}}_A$ and $\hat{\mathcal{V}}_B$, respectively. The Lagrangian reads

\begin{equation}
    \mathcal{L}=\sum\limits_{n}\Psi^\dagger_{n}(\mathbf{x}) (\hat{\mathcal{G}}^{-1}-\hat{\mathcal{V}}_A-\hat{\mathcal{V}}_B)\Psi_{n}(\mathbf{x}).
	\label{eq:Casimir_Lagrangian}
\end{equation}

\noindent Here, $\hat{\mathcal{G}}=(i\omega_n-\hat{H})^{-1}$ is the Matsubara Green’s function with fermionic Matsubara frequencies $\omega_n=(2n+1)\pi/\beta$, where $\beta$ is the inverse temperature. We adopt natural units with $k_B=\hbar=1$.  The partition function of the system is given by %coupled with $\hat{\mathcal{V}}_A$ and $\hat{\mathcal{V}}_B$
$\mathcal{Z}=\int\mathcal{D}\Psi^\dagger\mathcal{D}\Psi \exp(\int\mathcal{L} d\mathbf{x})$. The Casimir energy introduced by objects $A$ and $B$ is manifested by subtracting the free energy of the system in the absence of the local potentials, i.e., $\hat{\mathcal{V}}_A = \hat{\mathcal{V}}_B = 0$. Further subtracting the self-energy contributions (diagonal terms in the coordinate basis) yields the interaction energy between $A$ and $B$ \cite{Chiral_Casimir}

\begin{equation}
    E_c = -\frac{1}{\beta}\sum\limits_{n}\mathrm{ln}\,\mathrm{det}(1- T_A\mathcal{G}_{AB}  T_B\mathcal{G}_{BA}).
    \label{eq:Fermion_Casimir}
\end{equation}
	
\noindent Here, $T_A = \mathcal{V}_A (1-\mathcal{G}_{AA}\mathcal{V}_A)^{-1}$ and $T_B = \mathcal{V}_B (1-\mathcal{G}_{BB} \mathcal{V}_B)^{-1}$ are the scattering matrices of the two objects. $\mathcal{G}_{AA}$ ($\mathcal{G}_{BB}$) denotes the local Matsubara Green’s function at object $A$ ($B$), while $\mathcal{G}_{BA}$ ($\mathcal{G}_{AB}$) describes fermion propagation from $A$ ($B$) to $B$ ($A$).

\begin{figure}
    \centering
\includegraphics[height=3.2cm]{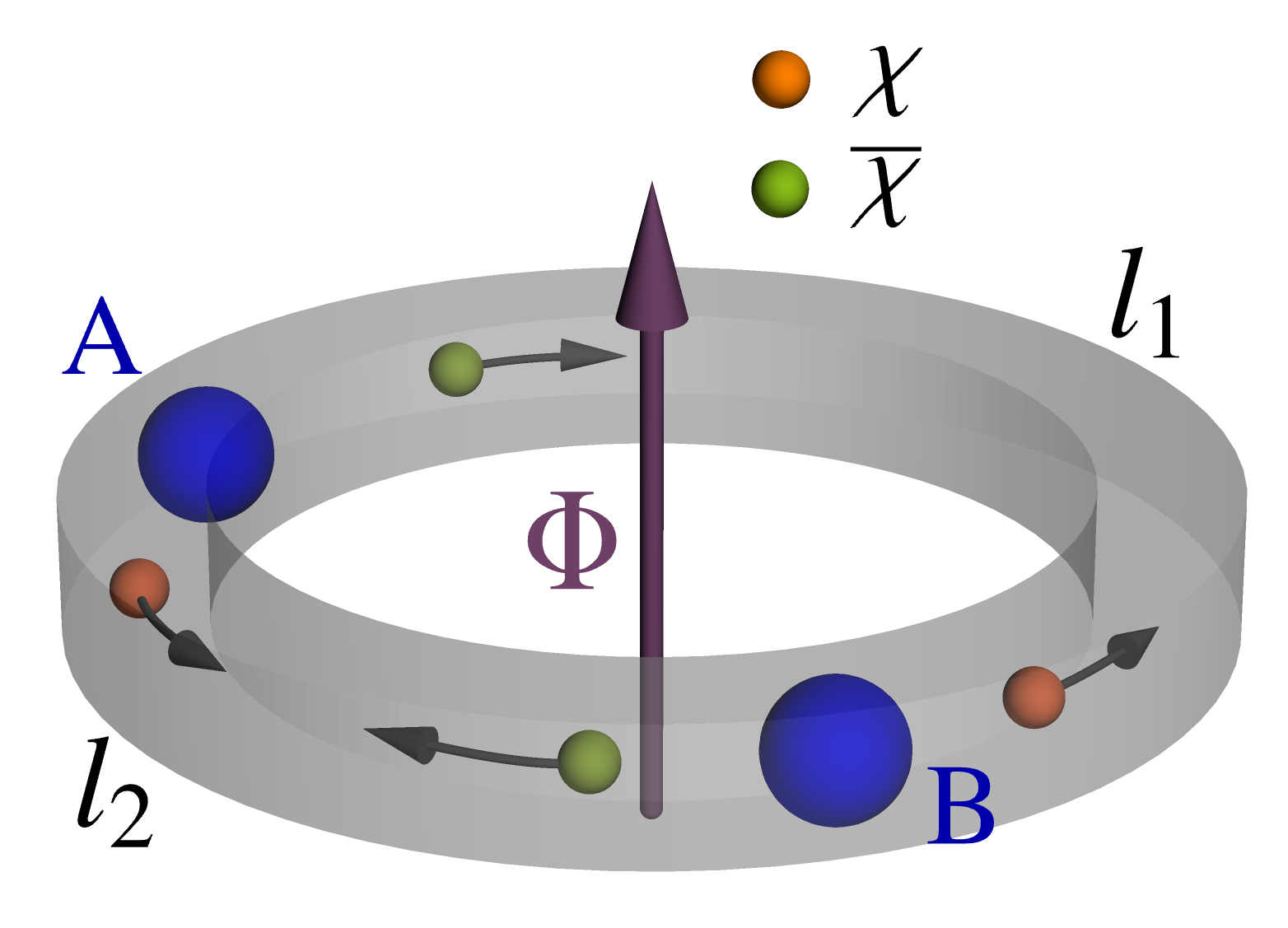}
    \caption{A mesoscopic ring supporting a persistent current and hosting two objects (localized states) $A$ and $B$. Spin-degenerate itinerant electrons with opposite chiralities $\chi$ and $\bar{\chi}$ propagate between $A$ and $B$ along two paths of lengths $l_1$ and $l_2$, respectively. %A magnetic flux $\Phi$.
    }
    \label{fig:ring}
\end{figure}

Previous studies of the Casimir effect \cite{Fermion_Casimir1, Chiral_Casimir} have primarily focused on bulk systems, in which $\mathcal{G}_{AB}$ and $\mathcal{G}_{BA}$ generally involve infinitely many propagation paths, rendering the interaction susceptible to decoherence among different paths. Moreover, controlled tuning of the Casimir effect in such geometries remains challenging, as it is largely determined by intrinsic material properties \cite{zhang2024magnetic}.
To circumvent these limitations, we consider a 1D setup in which the fermionic field resides on a mesoscopic ring, where $\Psi^{\dagger}_{\sigma}$ creates itinerant electrons with spin $\sigma$ that contribute to a persistent current \cite{buttiker1983josephson, persistent_current}. The spin-degenerate itinerant electrons on the ring are described by the Hamiltonian $\hat{H} = \sum_{\sigma=\uparrow,\downarrow}\Psi_\sigma^\dagger \left[ \frac{1}{2m^*} \left(\mathbf{p}+e\mathbf{A}\right)^2-\mu \right] \Psi_\sigma$, where $m^*$ is the effective mass of the electron. %Here, $m^*$ and $q$ denotes the effective electron mass and charge, respectively.
A magnetic flux $\Phi$ threading the ring enters via the Peierls substitution $\mathbf{p} \to \mathbf{p}+e\mathbf{A}$ with $\Phi = \oint_{l} d\mathbf{l} \cdot \mathbf{A}$, thereby providing a controllable tuning parameter.
We model objects $A$ and $B$ as two localized states at distinct positions on the ring. Electrons propagating between them are restricted to two 1D paths, corresponding to clockwise and counterclockwise motion (see Fig.~\ref{fig:ring}), which we label by opposite chiralities. In the chiral basis $\left( \ket{\chi}, \ket{\bar{\chi}} \right)^{\mathrm{T}}$, with $\chi = -\bar{\chi} = \pm 1$, the Green’s functions $\mathcal{G}_{BA}$ and $\mathcal{G}_{AB}$ take the form

\begin{equation}
    	\mathcal{G}_{BA}=
    	im^*\begin{pmatrix}
    		\frac{e^{\pm i\kappa l_{1}}}{\kappa}e^{ie\Phi_{1}} & 0\\
    		0 & \frac{e^{\pm i\kappa l_{2}}}{\kappa} e^{-ie\Phi_{2}}
    	\end{pmatrix},
    \label{eq:Green_GAB}
\end{equation}

\noindent and

\begin{equation}
    	\mathcal{G}_{AB}=
    	im^*\begin{pmatrix}
    		\frac{e^{\pm i\kappa l_{2}}}{\kappa}e^{ie\Phi_{2}} & 0\\
    		0 & \frac{e^{\pm i\kappa l_{1}}}{\kappa} e^{-ie\Phi_{1}}
    	\end{pmatrix},
    \label{eq:Green_GBA}
\end{equation}

\noindent respectively, for both the $\sigma=\ \uparrow$ and $\sigma=\ \downarrow$ itinerant electrons. Here, $\kappa=\sqrt{2m^*(i\omega_n+\mu)}$ is the electron momentum with $\pm$ corresponds to $n\geq0$ and $n<0$, respectively. $\mu$ is the chemical potential, and $l_1$ ($l_2$) indicates the length of the upper (lower) half of the ring (see Fig. \ref{fig:ring}). Correspondingly, $e\Phi_j = e\int_{l_j} d\mathbf{l} \cdot \mathbf{A}$ is the geometric phase accumulated via the Aharonov–Bohm (AB) effect ($j=1,2$).

The itinerant electrons are scattered by the local potentials $\hat{\mathcal{V}}_A$ and $\hat{\mathcal{V}}_B$, which can reverse their chirality. In the chiral basis, the scattering matrices $T_A$ and $T_B$ take the general form

\begin{equation}
T_A=T_B=t\tau_0+r\tau_x,
    \label{eq:matrix}
\end{equation}

\noindent where $\tau_i$ denote the Pauli matrices in the chiral basis. The transmission and reflection amplitudes $t$ and $r$ satisfy the particle-number conservation condition $|t|^2+|r|^2=1$. For simplicity, hereafter we assume identical scatterers $A$ and $B$, such that $T_A = T_B$.
Two elementary processes contribute to the fermion-mediated Casimir interaction on the mesoscopic ring: (i) electrons traversing closed loops around the ring and enclosing the magnetic flux, and (ii) electrons undergoing backscattering from objects $A$ or $B$ and returning to their original positions. The former process is associated with the AB effect, such that the corresponding AB phase provides a tunable control of the Casimir energy (see Supplemental Material \cite{SupplementaryMaterials}). We therefore term this flux-tunable fermion-mediated Casimir interaction, arising from its interplay with the AB effect, the Aharonov–Bohm–Casimir (ABC) effect.

%%%%%%%%%% THIS IS A CUTTING LINE %%%%%%%%%%

\textit{Spin-$1/2$ qubits.}---We now consider a concrete realization in which objects $A$ and $B$ are localized spin-$1/2$ moments serving as qubits that encode quantum information in their spin polarization \cite{DanielLoss_spin_qubit}. Such a setup can be implemented, for example, using two QDs, each hosting a single spin-$1/2$ electron, embedded on a mesoscopic ring. The system remains described by Eq.~(\ref{eq:Casimir_Lagrangian}) under the substitution $\hat{\mathcal{V}}_{A(B)} \rightarrow (1/2)J_{A(B)} \mathbf{S}_{A(B)} \cdot \boldsymbol{\sigma} \delta(\mathbf{x}-\mathbf{x}_{A(B)})$, where $\mathbf{S}_A$ and $\mathbf{S}_B$ denote the localized spins, coupled to the itinerant electron spin $\boldsymbol{\sigma}$ with exchange couplings $J_A$ and $J_B$, respectively. In the weak-coupling regime, expanding Eq.~(\ref{eq:Fermion_Casimir}) to leading order yields an effective interaction Hamiltonian between $\mathbf{S}_A$ and $\mathbf{S}_B$,

\begin{equation}
    \begin{split}
        \hat{H}_{AB} =& \frac{1}{\beta}\sum\limits_{n}\mathrm{Tr}[(J_AT_A) \mathbf{S}_A \cdot \boldsymbol{\sigma}\mathcal{G}_{AB} (J_BT_B) \mathbf{S}_B \cdot \boldsymbol{\sigma}\mathcal{G}_{BA}]\\
        =& \frac{4}{\beta}J_AJ_B\sum\limits_{n} \mathrm{Tr}(T_A\mathcal{G}_{AB}  T_B\mathcal{G}_{BA}) \mathbf{S}_A \cdot \mathbf{S}_B.
    \end{split}
    \label{eq:lowest_order}
    \end{equation}

As discussed above, the general form of the scattering matrices [Eq.~(\ref{eq:matrix})] indicates that two elementary processes contribute to the Casimir interaction. Accordingly, the effective interaction Hamiltonian $\hat{H}_{AB}$ reads
%\noindent Inserting the general form of the scattering matrices mentioned in Eq. (\ref{eq:matrix}), the effective Hamiltonian describing the interaction between two localized spin $\mathbf{S}_A$ and $\mathbf{S}_B$ can be rewritten in the form of

\begin{equation}
    \hat{H}_{AB} = -\left( J_t \cos\theta + J_r\right) \mathbf{S}_A \cdot \mathbf{S}_B.
    \label{eq:RKKY}
\end{equation}
%J\left(\theta,l_1,l_2\right) \mathbf{S}_A \cdot \mathbf{S}_B

\noindent where we define the dimensionless phase $\theta \equiv 2\pi\Phi/\Phi_0$, with $\Phi=\Phi_1+\Phi_2$ the magnetic flux threading the ring and $\Phi_0 = h/e$ the flux quantum.
The term $J_t$, proportional to $|t|^2$, corresponds to electron traversal around the entire ring and depends on the ring circumference $l_1+l_2$ as

\begin{equation}
    J_t = 4|t|^2m^{*2}J_AJ_B\frac{1}{\beta}\sum\limits_{n\geq0}\frac{e^{i\kappa\left(l_1+l_2\right)}}{\kappa^2} + \mathrm{h.c.},
    \label{eq:trans}
\end{equation}
	
\noindent while the term $J_r \propto |r|^2$ corresponds to backscattering processes and depends on the lengths of the two paths, $l_1$ and $l_2$, as
	
\begin{equation}
    J_r = 2|r|^2m^{*2}J_AJ_B\frac{1}{\beta}\sum\limits_{n\geq0}\frac{e^{2i\kappa l_1}+e^{2i\kappa l_2}}{\kappa^2} + \mathrm{h.c.}.
    \label{eq:reflects}
\end{equation}
	
\begin{figure}
    \centering
    \begin{minipage}[t]{9cm}
        \centering
        \includegraphics[height=3.5cm]{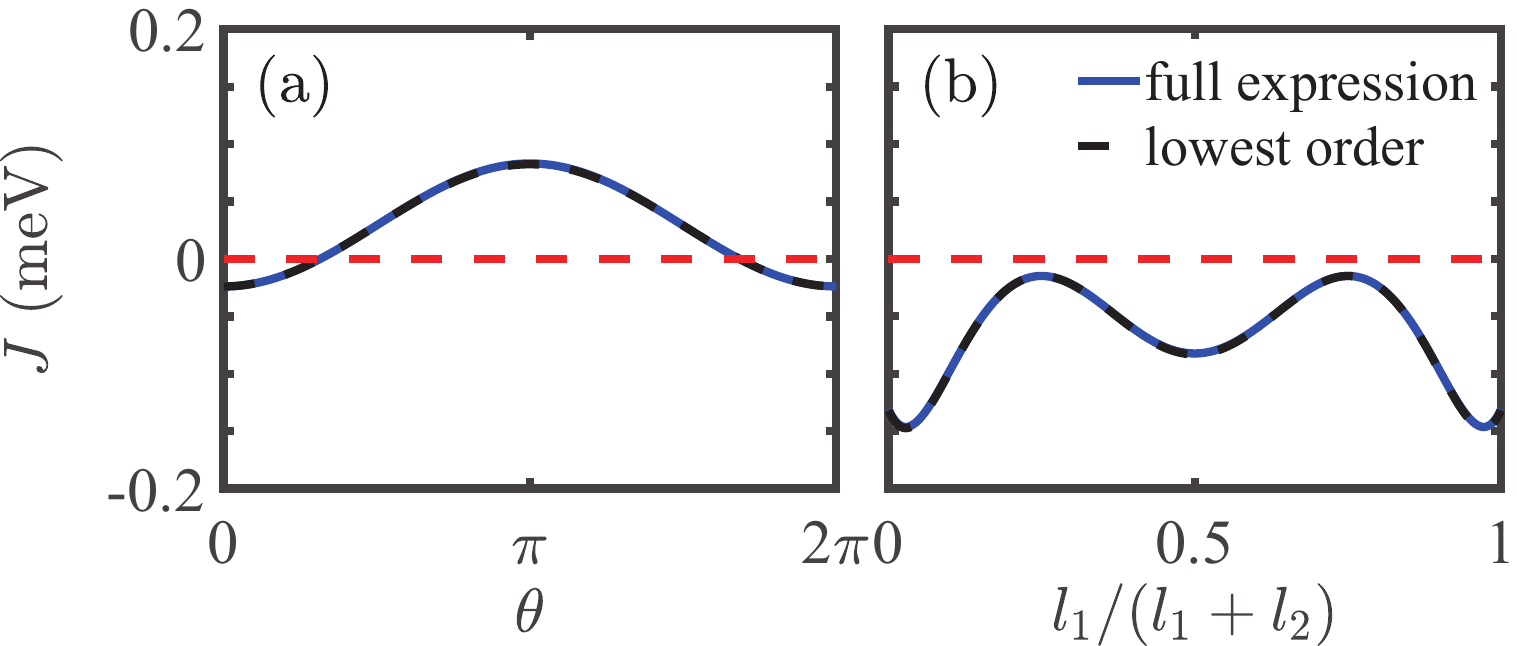}
    \end{minipage}
    \caption{The ABC interaction strength in its lowest-order expansion (dashed black line) and in the full expression (solid blue line). 
    (a) ABC interaction strength as a function of $\theta$. Here, $l_1 = 10\mathrm{nm}$, $l_2 = 40\mathrm{nm}$, $r=3/5$, $t=4/5$, $m^*=0.067m_e$ \cite{exppersistent} ($m_e$ is free electron mass),  $\mu=10$meV, and $T=0.1 \mathrm{K}$. The Matsubara frequencies are summed up to $n = 1000$.
    (b) ABC interaction strength as a function of $l_1/(l_1+l_2)$. Here, $l_1+l_2 = 50\mathrm{nm}$, and $\theta=0$, with all other parameters the same as those in (a).}
    \label{fig:Casimir_flux}
\end{figure}

As shown in Eq.~(\ref{eq:RKKY}), the electron traversal process contributing to the fermion-mediated Casimir interaction depends on the magnetic flux $\Phi$, exhibiting a periodicity set by the flux quantum $\Phi_0$. This periodic modulation is confirmed by numerical calculations, as shown in Fig.~\ref{fig:Casimir_flux}(a). The flux-induced sign reversal of the interaction follows from the generalized theorem for fermion-mediated Casimir interactions \cite{Fermion_Casimir1}, whereby the magnetic flux effectively reverses the chiral symmetry of the Green’s function (see Supplemental Material \cite{SupplementaryMaterials}).

% Otherwise, when itinerant electrons enclose magnetic flux, the sign of interaction is determined by $-\mathrm{sgn}(\cos\left[ k_F\left(l_1+l_2\right) \right])$ owing to the same chirality of the Green's function. However, the external magnetic flux reverses such chirality and flips the sign of the interaction \cite{Fermion_Casimir1} (see Supplemental Material \cite{SupplementaryMaterials}).

One may notice that the lowest-order expression of the ABC effect resembles the Ruderman–Kittel–Kasuya–Yosida (RKKY) interaction \cite{RKKY1, RKKY2, RKKY3, RKKY4, ABRKKY} between magnetic moments. In conventional settings, the RKKY interaction is mediated by electrons in $d$-dimensional bulk materials (typically $d=3$), leading to a power-law decay $\sim l^{-d}$ with $l$ the distance between the two moments. Moreover, in bulk systems the presence of an external magnetic field introduces a magnetic length scale and modifies electron propagation through orbital quantization effects, leading to dephasing and suppression of interference contributions \cite{weak_localization_suppress_inteference, Chakravarty_1986_weak_localization} rather than a simple periodic flux dependence.
In contrast, the ABC effect exhibits periodic oscillations with respect to magnetic flux as well as a long-range interaction scaling as $l^{-1}$, with  $J_t = 4C|t|^2\frac{\cos\left[ k_F\left(l_1+l_2\right) \right]}{k_F(l_1+l_2)}$ and $J_r = 2C|r|^2[\frac{\cos\left(2k_Fl_1\right)}{2k_Fl_1} + \frac{\cos\left(2k_Fl_2\right)}{2k_Fl_2}]$ in the zero temperature and long-wavelength limit $k_F(l_1+l_2)\ll1$, where $C=\frac{m^{*}J_AJ_B}{\pi}$ and $k_F$ is the Fermi momentum of the itinerant electrons in the mesoscopic ring. Compared with the exponentially decaying Heisenberg exchange interaction \cite{rkky-exp}, this behavior constitutes a genuine long-range interaction.

\begin{figure*}
	
	\begin{minipage}[t]{0.3\textwidth}
		\centering
		\includegraphics[width=1.2
		\linewidth]{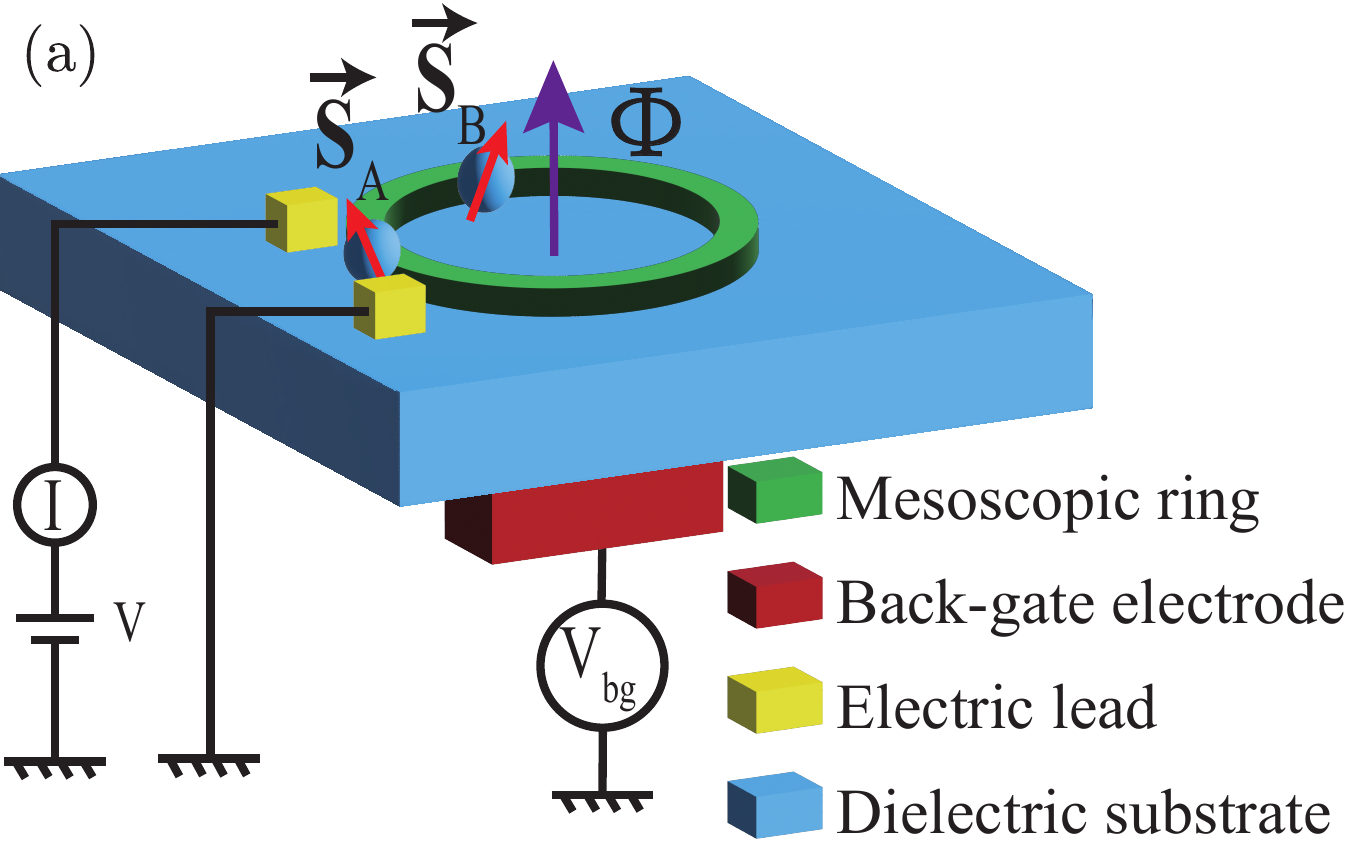}
		
	\end{minipage}
	\hfill
	\begin{minipage}[t]{0.3\textwidth}
		\centering
		\includegraphics[width=\linewidth]{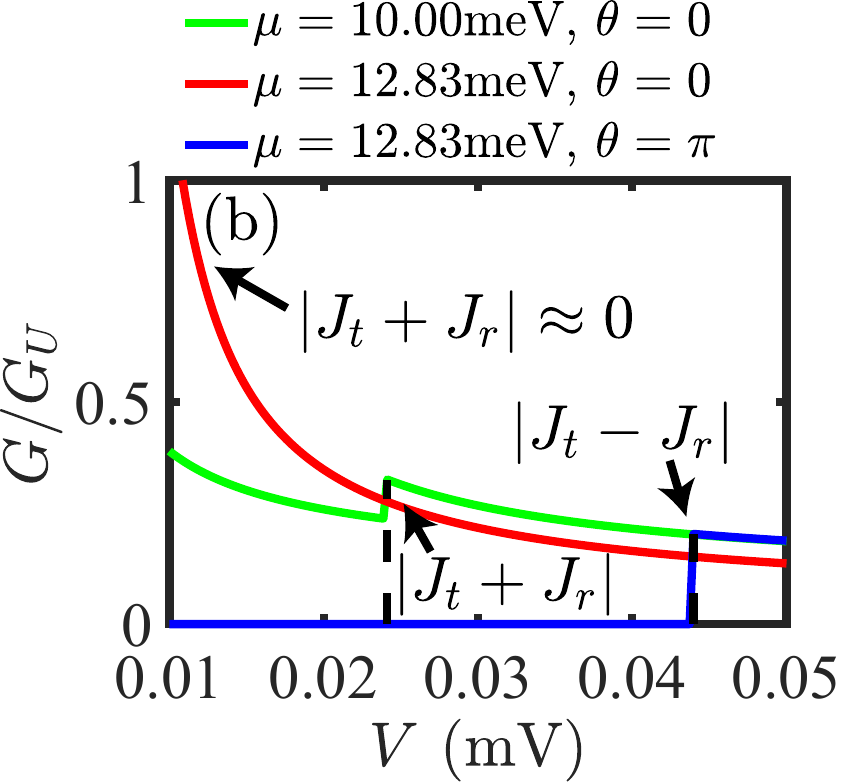}
		
	\end{minipage}
	\hfill
	\begin{minipage}[t]{0.3\textwidth}
		\centering
		\includegraphics[width=\linewidth]{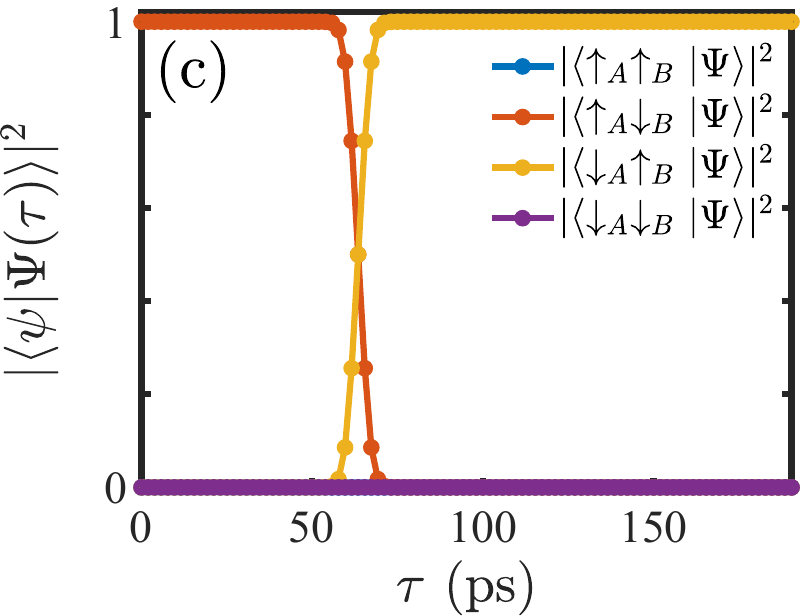}
	\end{minipage}
	
	\caption{
		(a) Schematic of the device for measuring the differential conductance and implementing the $\mathrm{SWAP}^\alpha$ gate.
		(b) Differential conductance $G$ (in the unit of $G_U \equiv 2e^2/h$) for different interaction strength at Kondo temperature $T_k=0.01$K.
		Green and red curves: in the absence of the magnetic flux ($\theta=0$), the conductance exhibits peaks at different bias voltages for different chemical potentials $\mu$. 
		Blue curve: the peak position reaches its maximum at $\theta=\pi$. 
		Other parameters are the same as those in Fig. \ref{fig:Casimir_flux}. 
		(c) Time evolution of a qubit initialized in the state $\ket{\Psi(\tau=0)} = \ket{\uparrow_A\downarrow_B}$ under a magnetic-flux pulse $\theta(\tau) = \theta_0 \mathrm{sech}^2[2\lambda(\tau-\tau_0)]$ with $\tau_0=5/\lambda$, $\theta_0=\pi$, and $2\lambda=0.2279(\mathrm{ps})^{-1}$. Other parameters are the same as those for the red curve in Fig.~\ref{fig:scheme}(b).
	}
	\label{fig:scheme}
\end{figure*}

In addition, as shown in Fig.~\ref{fig:Casimir_flux}, higher-order contributions to the ABC interaction, corresponding to multiple traversals of the ring and/or repeated backscattering events, do not significantly modify the results in the weak-coupling regime. This indicates that the ABC effect is dominated by two elementary processes, namely single traversal around the ring and single backscattering, rather than higher-order multiple-round or multiple-scattering events. Consequently, the ABC effect only requires a mesoscopic ring with a coherence length comparable to its circumference, consistent with the conditions under which persistent currents have been experimentally observed \cite{PhysRevLett.64.2074, PhysRevLett.67.3578, persistent_current, persistent_current_exp_2009}.

%%%%%%%%%% THIS IS A CUTTING LINE %%%%%%%%%%

\textit{Controlled $\mathrm{SWAP}^\alpha$ gates.}---The ABC effect on a mesoscopic ring is tunable via the applied magnetic flux, thereby enabling controlled manipulation of local magnetic moments. In particular, the unitary evolution $\hat{U} = \exp(-i\int_{\tau_{s}}^{\tau_{e}} \mathrm{d}\tau \hat{H}_{AB})$ \cite{time-AB1, time-AB2, time-AB3} generated by the ABC interaction $\hat{H}_{AB}$ in Eq.~(\ref{eq:RKKY}) over the time interval $[\tau_s,\tau_e]$ implements a two-qubit quantum gate (up to a global phase) \cite{sqrtSWAP_universal_quantum_gate_3}

\begin{equation}
    \hat{U} = \begin{pmatrix}
        1&0&0&0\\
        0&\frac{1+e^{i\pi\alpha}}{2}&\frac{1-e^{i\pi\alpha}}{2}&0\\
        0&\frac{1-e^{i\pi\alpha}}{2}&\frac{1+e^{i\pi\alpha}}{2}&0\\
        0&0&0&1
    \end{pmatrix}
    \label{eq:SWAP}
\end{equation}

\noindent acting on the two qubits encoded in the spin polarizations of $\mathbf{S}_A$ and $\mathbf{S}_B$ as $\left( \Ket{\uparrow_A\uparrow_B}, \Ket{\uparrow_A\downarrow_B}, \Ket{\downarrow_A\uparrow_B}, \Ket{\downarrow_A\downarrow_B} \right)^{\mathrm{T}}$.
Equation~(\ref{eq:SWAP}) corresponds to the $\mathrm{SWAP}^\alpha$ gate \cite{SWAP1}, where the parameter $\alpha$ is determined by $\alpha\pi = 4\int_{\tau_s}^{\tau_{e}} \mathrm{d}\tau \left(J_t \cos\theta + J_r\right)$. For $\alpha=1$, the operation reduces to the $\mathrm{SWAP}$ gate, which exchanges the quantum states of $\mathbf{S}_A$ and $\mathbf{S}_B$. For non-integer $\alpha$ (e.g., $\alpha=1/2$, corresponding to the $\sqrt{\mathrm{SWAP}}$ gate), the operation realizes a non-Clifford $\mathrm{SWAP}^\alpha$ gate capable of generating entanglement, which is essential for universal quantum computation \cite{sqrtSWAP_universal_quantum_gate_1, sqrtSWAP_universal_quantum_gate_2}.	

Precise implementation of the $\mathrm{SWAP}^{\alpha}$ gate requires accurate tuning of the  $\alpha = \frac{4}{\pi} \int_{\tau_s}^{\tau_{e}} \mathrm{d}\tau \left(J_t \cos\theta + J_r\right)$. The Casimir interaction strengths $J_t$ and $J_r$ depend on both the Fermi momentum $k_F$ and microscopic interaction details, including the transmission and reflection amplitudes $t$ and $r$, as well as the exchange couplings $J_A$ and $J_B$. 
While $k_F$ can be tuned experimentally via the electron density in the mesoscopic ring, the interaction parameters are sample-dependent and generally not known a priori. Consequently, device-specific calibration of $J_t$ and $J_r$ is required. Fortunately, they can be extracted from differential conductance measurements exploiting the Kondo effect \cite{Kondo1, Kondo2, ABkondo,RKKY1supp1}.

%However, the persistent current induced by such magnetic field also generates a self-established magnetic flux that would destabilize the system, which imposes $\theta=0$ as well as $J_t +J_r=0$ after the gate operation ($t\geq \tau_e$). 

As illustrated by the device shown in Fig.~\ref{fig:scheme}(a), the differential conductance is measured using two metallic leads coupled to the QD hosting spin $\mathbf{S}_A$. Owing to the Kondo effect \cite{Kondo1, Kondo2, ABkondo}, the differential conductance exhibits a peak when the bias voltage $V$ satisfies $eV = |J_t \cos\theta + J_r|$. As the magnetic flux $\theta$ is varied, the peak position oscillates periodically within the range from $eV =|J_t-J_r|$ to $eV=|J_t+J_r|$. The values of $J_t$ and $J_r$ can thus be extracted from the extrema of the peak positions.

%To realize a desired $\mathrm{SWAP}^\alpha$ gate, such a device requires that the two QDs are decoupled in absence of magnetic flux {\color{blue} and maintain flux controllability, which implies that $|J_t +J_r|=0$ and $|J_t-J_r|\neq 0$ at the same time. However, those conditions are satisfied only if $l_1\neq l_2$; therefore, such a quantum gate device requires that the two path distances between two QDs on the mesoscopic ring be unequal.}

Based on the above discussion, our scheme for implementing the $\mathrm{SWAP}^\alpha$ gate proceeds in three steps:

(i) \textit{Tuning the ABC interaction to zero.} We first set the magnetic flux to zero ($\theta = 0$) and tune the back-gate voltage [see Fig.~\ref{fig:scheme}(a)] to adjust the chemical potential $\mu$, thereby controlling the Fermi momentum $k_F$. For a generic $\mu$ [green curve in Fig.~\ref{fig:scheme}(b)], the differential conductance exhibits a peak at a finite bias, $eV = |J_t + J_r| \neq 0$. By monitoring the peak position while tuning $\mu$, we identify the condition where the peak shifts to $V \approx 0$ [red curve in Fig.~\ref{fig:scheme}(b)], indicating $J_t + J_r = 0$ and hence a vanishing ABC interaction. This condition can be achieved provided that $l_1 \neq l_2$ (see Supplemental Material \cite{SupplementaryMaterials}). The two qubits are thus decoupled, and no time evolution occurs unless a nonzero magnetic flux is applied.

(ii) \textit{Extracting $J_t$ and $J_r$.} Keeping the back-gate voltage fixed at the value determined above, we apply a finite magnetic flux $\theta$. 
%As shown in the blue curve in Fig.~\ref{fig:scheme}(b), an additional conductance peak arising from the Kondo resonance of a single localized spin-$1/2$ moment  is present near $V=0$ in ferromagnetic region. In addition, the 
The position of the conductance peak associated with $eV = |J_t \cos\theta + J_r|$ oscillates periodically within the range $eV \in [|J_t + J_r|, |J_t - J_r|]$ (see Supplemental Material \cite{SupplementaryMaterials}). By varying $\theta$, the values of $J_t$ and $J_r$ can be extracted from the extrema of the peak positions, occurring at $\theta = 0$ and $\theta = \pi$ [red and blue curves in Fig.~\ref{fig:scheme}(b)]. 

(iii) \textit{Implementing the $\mathrm{SWAP}^\alpha$ gate.} With $J_t$ and $J_r$ calibrated, a $\mathrm{SWAP}^\alpha$ gate is realized by applying an appropriate time-dependent magnetic flux $\theta(\tau)$ such that $4\int_{\tau_{s}}^{\tau_{e}} \mathrm{d}\tau \left[J_t \cos\theta(\tau) + J_r\right] = \alpha\pi$. A convenient choice is a pulse of the form $\theta(\tau) = \theta_0 \mathrm{sech}^2[2\lambda(\tau-\tau_0)]$ \cite{revolution1, SWAP1, revolution2}, where $\tau_0$ is time offset, $\theta_0$ and $\lambda$ denote the pulse amplitude and inverse width, respectively. As shown in Fig.~\ref{fig:scheme}(c), an initial state $\ket{\Psi(\tau=0)} = \ket{\uparrow_A\downarrow_B}$ evolves into $\ket{\Psi(\tau \to \infty)} = \ket{\downarrow_A\uparrow_B}$ under such a pulse, demonstrating the implementation of a $\mathrm{SWAP}$ gate ($\alpha = 1$). More generally, non-Clifford $\mathrm{SWAP}^\alpha$ gates can be realized by appropriately choosing $\theta_0$ and $\lambda$.

%Therefore, as the red and black line shown in Fig. \ref{fig:diffcon}, upon varying the applied back-gate voltage to control the electron density on the mesoscopic ring, the bias voltage $V$ corresponding to the conductance peak at $eV=|J_t+J_r|$ is expected to shift and vanish at $V=0$. Then, by introducing the magnetic flux $\theta = \Phi/\Phi_0$, such Kondo peak oscillate periodically between $eV=|J_t+J_r|$ and $eV=|J_t-J_r|$ (see red and blue line in Fig.\ref{fig:diffcon}). Consequently, the values of $J_t$ and $J_r$ can be extracted from the positions of the conductance peaks.
	
%Noticing that the persistent current in the mesoscopic ring itself will also induce a built-in magnetic flux. The presence of this flux tends to minimize the energy of the mesoscopic ring \cite{persistent_current}. Therefore, the equilibrium persistent current is expected to generate a built-in magnetic flux $\theta_0$, which makes the interaction between $\mathbf{S}_A$ and $\mathbf{S}_B$ vanishes as $J_t \cos\theta_0 + J_r = 0$. In other words, the two qubits do not evolve until an external magnetic flux is applied.

\textit{Discussion.}---The conventional Casimir effect is mediated by photons, whose coherence length is far beyond the distance between mirrors \cite{photon_coherence_length}. In contrast, electrons in solid-state systems typically exhibit much shorter coherence lengths. However, as discussed above, the ABC effect only requires the coherence length of itinerant electrons to be comparable to the circumference of the mesoscopic ring. This condition is the same as that required for observing persistent currents, which have been extensively studied over past decades \cite{PhysRevLett.64.2074, PhysRevLett.67.3578, persistent_current, persistentcurrent1, persistentcurrent2,exppersistent}. Moreover, persistent currents can be derived within the same formalism as fermion-mediated Casimir interactions (see Supplemental Material \cite{SupplementaryMaterials}). These considerations support the experimental feasibility of observing the ABC effect in mesoscopic ring systems.

%{\color{blue} In the previous works, it has been proved that a spin-$1/2$ quantum dot (QD) could be realized by enhancement of the on-site Coulomb energy \cite{Kondo1}. Meanwhile, the two possible states in spin-$1/2$ system was already used to represent the qubits in quantum computing and had  been already achieved in the spin-$1/2$ QDs system with Heisenberg interaction \cite{SWAP2}. Therefore, by introducing spin-$1/2$ QDs on the system, it can thus enable the implementation of the quantum gate operation.} %此段为背景介绍，先放着。

Although higher-order terms in the ABC effect [Eq.~(\ref{eq:Fermion_Casimir})] are subleading in the weak-coupling regime, they nonetheless introduce a finite source of gate error. In the Supplemental Material \cite{SupplementaryMaterials}, we analyze second-order contributions, such as those proportional to $\cos 2\theta$, and find that they lead to an infidelity of the $\mathrm{SWAP}^{\alpha}$ gate of approximately $0.9\%$. This level of two-qubit gate error is comparable to that achieved in state-of-the-art superconducting transmon circuits \cite{Google_Willow} and neutral-atom platforms \cite{Lukin}, showing that the ABC effect provides a viable route to high-fidelity, non-Clifford two-qubit gate operations. Finally, by introducing additional interactions such as strong spin–orbit coupling \cite{SOC1} into the Hamiltonian, anisotropic exchange terms (for example, Ising or Dzyaloshinskii–Moriya interactions) may become dominant, enabling the realization of other flux-controlled quantum gates such as the Controlled-Z gate.

\textit{Acknowledgments.}---We thank Jian-Hui Zhou for helpful discussions. This work is financially supported by the National Key R$\&$D Program of China (Grant No. 2024YFA1409000), Quantum Science and Technology-National Science and Technology Major Project (Grant No. 2021ZD0302400), the National Natural Science Foundation of China (Grant No. 12304194, No. 12574171, and No. 12374332), Innovation Program for Quantum Science and Technology Grant No. 2021ZD0301900, Shanghai Municipal Science and Technology (Grant No. 24DP2600100), Shanghai Pilot Program for Basic Research - Fudan University 21TQ1400100 (25TQ003), and Shanghai Science and Technology Innovation Action Plan (Grant No. 24LZ1400800).

\bibliography{Casimir_SWAP} 
\end{document}